\documentclass[11pt,twoside]{article} 
\usepackage{./asp2014} 
 
\aspSuppressVolSlug 
\resetcounters 
 
\begin{document} 

\title{The giant radio flare of Cygnus X-3 in September  2016}
\author{Trushkin S.A.,$^{1,2}$, Nizhelskij N.A.,$^1$, Tsybulev P.G.$^1$ and Zhekanis G.V.$^1$ }
\affil{$^1$Special astrophysical Observatory RAS, Nizhnij Arkhys, Karachaevo-Cherkassia, Russia; \email{satr@sao.ru}}
\affil{$^2$Kazan Federal University, Kazan, Republic of Tatarstan, Russia}

\paperauthor{Trushkin S.A.}  {satr@sao.ru} {}{Special astrophysical Observatory RAS}{Radio astophysics laboratory}{Nizhnij Arkhyz}{KChR}{369167}{Russia}
\paperauthor{Nizhelskij N.A.}{nizh@sao.ru} {}{Special astrophysical Observatory RAS}{Radio continuum laboratory}{Nizhnij Arkhyz}{KChR}{369167}{Russia}
\paperauthor{Tsybulev P.G.}  {peter@sao.ru}{}{Special astrophysical Observatory RAS}{Radio continuum laboratory}{Nizhnij Arkhyz}{KChR}{369167}{Russia}
\paperauthor{Zhekanis G.V.}  {gvz@sao.ru}  {}{Special astrophysical Observatory RAS}{RATAN adjustment group}{Nizhnij Arkhyz}{KChR}{369167}{Russia}

\begin{abstract}
In the long-term multi-frequency monitoring program of the microquasars
with RATAN-600 we discovered the giant flare from X-ray binary Cyg~X-3
on 13 September 2016. It happened after 2000 days of the 'quiescent
state' of the source passed after the former giant flare ($\sim18$ Jy) in
March 2011.  We have found that during this quiet period the  hard X-ray
flux (Swift/BAT, 15-50 keV) and radio flux (RATAN-600, 11 GHz) have been
strongly anti-correlated. Both radio flares occurred after transitions of
the microquasar to a 'hypersoft' X-ray state that occurred in February 2011
and in the end of August 2016. The giant flare was predicted by us in the
first ATel (\cite{T3}). Indeed after dramatic decrease of the hard X-ray
Swift 15-50 keV flux and RATAN 4-11 GHz fluxes (a 'quenched state')
a small flare (0.7 Jy at 4-11 GHz) developed on MJD 57632
and then on MJD 57644.5 almost simultaneously with X-rays radio flux rose
from 0.01 to 15 Jy at 4.6 GHz during few days. The rise of the flaring flux
is well fitted by a exponential law that could be a initial phase of
the relativistic electrons generation by internal shock waves in the jets.
Initially spectra were optically thick at frequencies lower 2 GHz and
optically thin at frequencies higher 8 GHz with typical spectral index
about $-0.5$. After maximum of the flare  radio fluxes at all
frequencies faded out with exponential law.
\end{abstract}

\subsection*{Monitoring program of microquasars with the RATAN-600 telescope}

We have carried out the long-term monitoring almost daily measurements
during a year  GRS1915+105, SS433, Cyg~X-1, Cyg~X-3, LSI+61d303, LS5039
with RATAN-600 at 2.3, 4.6, 8.2, 11.2, 21.7 GHz (sometimes at 1.2 and
30 GHz) during last 6 years (or more 2000 days) (\cite{T1,T2}).  We have
detected a lot of very bright flares (more than 1.5 Jy at 4.8 GHz) from
SS433. Often, once per month we detected the bright (>100 mJy) flares
from GRS1915+105. We have detected persistent but variable radio emission
from Cyg~X-1. Already during 36 orbital periods (26.5d) we continue to
study the super-orbital modulation ($P_2=1666$days) of the flaring radio
emission from LSI+61d303. The mean per some orbits  radio light curves
depend  strongly on phase $P_2$.  Almost 2000 days of the 'quiescent
state' of the Cyg~X-3  have passed after the former giant flare ($\sim
18$ Jy) in the end of March 2011.  We have detected it with RATAN-600
at 2.3-30 GHz.  By the way we have found that during this quiet period
the hard X-ray flux (Swift/BAT, 15-50 keV) and radio flux (RATAN-600,
11 GHz) were strongly and  anti-correlated ($\rho = -0.85$) (\ref{fig1}).
The nature of this linear regression could be related with properties
of the compact radio jets, forming during such 'quiescent' state and
strongly depending on an accretion rate on to a black hole (or a neutron star).

\articlefigure{fig1_lc_2011-16.eps}{fig1}{%
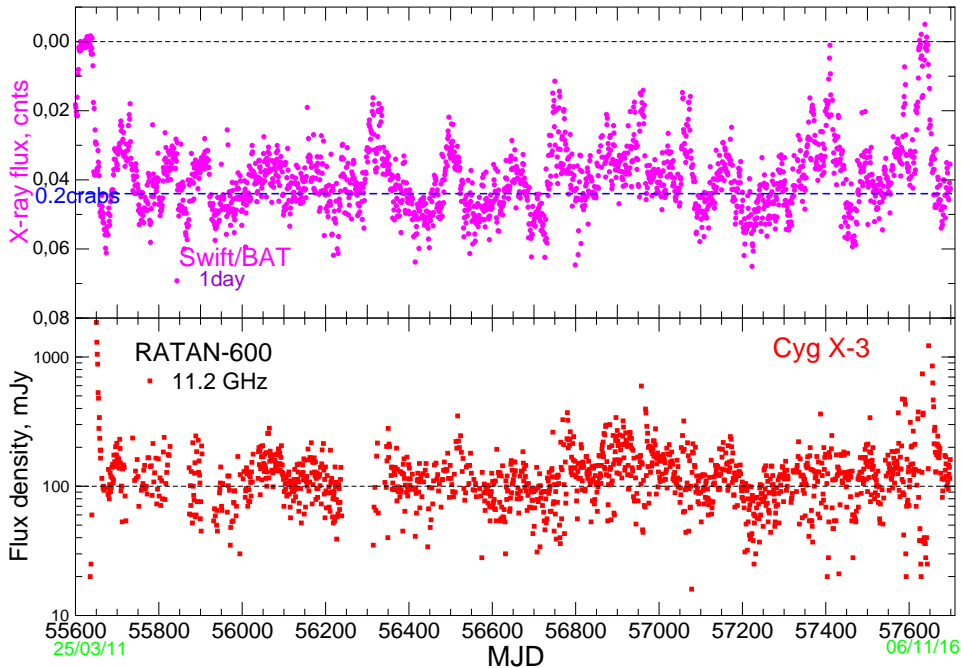
The light curves of Cyg~X-3 at 11 GHz (RATAN-600)
and at 15-50 keV (Swift/BAT) during 2011-2016.
For the best comparison the axis of X-ray fluxes is directed downwards.}
 
\subsection*{A new flare of Cyg~X-3 in September 2016}

The last two giant radio flares occurred  after transitions of the
microquasar to 'hypersoft' X-ray state that occurred in February 2011 and
in the end of August 2016.  The flux of the recent flare on MJD 57643.8
rose from 0.01 to 15 Jy at 4.6 GHz during five days. The giant flare was
predicted by us (\cite{T3}). Indeed after a dramatic decrease
of the hard X-ray 15-50 keV flux (Swift/BAT) lower the detection level
radio fluxes decreased also to 10-30 mJy (a 'quenched state'). But
then the both band fluxes increased in a small flare, about 0.7 Jy
at 4-11 GHz, then fade to a quenched state again, and then at last
increased almost simultaneously with increase of the hard X-ray flux.
The rising of the flaring flux is well fitted by a exponential law
$\propto \exp[(t-t_0)/0.54\rm{day}]$ at 11 GHz, where $t_0 = 57644.5$
is a probable start MJD-date of the flare. After maximum of the flare
the all radio fluxes at all frequencies faded out with exponential
law $\propto \exp[- (t-t_m)/2\rm{day}]$ where $t_m = 57650.7$ is the
date of the maximal fluxes and  Cyg~X-3 came back to 'quiescent state'
on 18 October 2016 \ref{fig2}.
After three-five days the flaring spectra became optically thin at 11-22 GHz
with same spectral index \ref{fig3}. Its value equal to $-0.5$
($S_{\nu} \propto \nu^{\alpha}$ is  direct evidence of the
first-order Fermi acceleration, although the evolved
internal shock is probably relativistic.

\articlefigure{fig2_lc_rat.eps}{fig2}{%
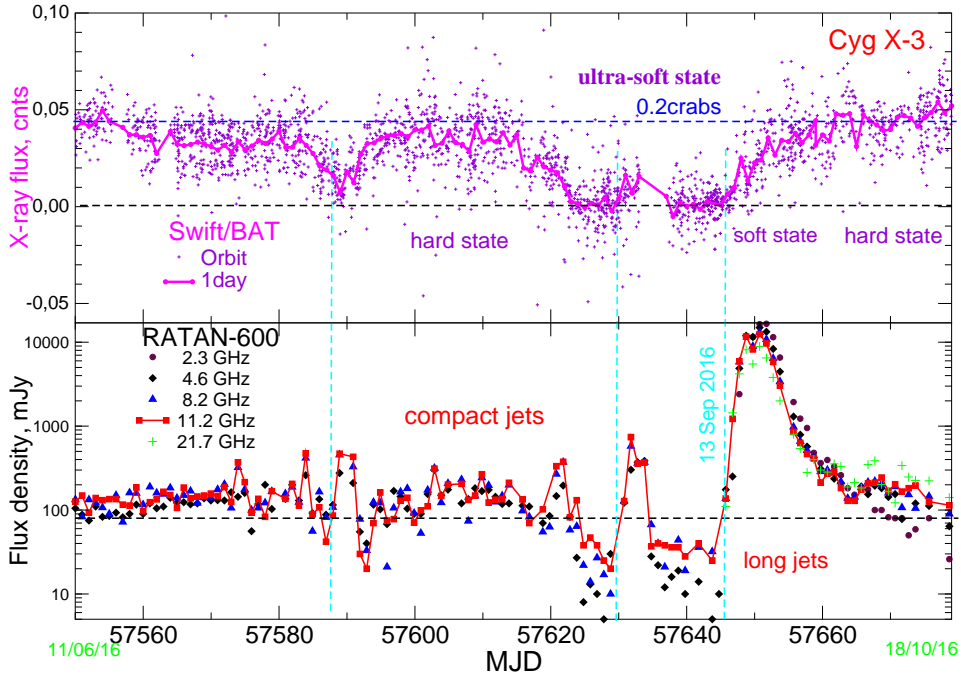
Light curves before or during the flare at X-ray 15-50 KeV (top)
and the multi-frequency data of the RATAN measurements (below).
Characteristic X-ray states of the binary are marked.}

\articlefigure{fig3_spectra_10d.eps}{fig3}{%
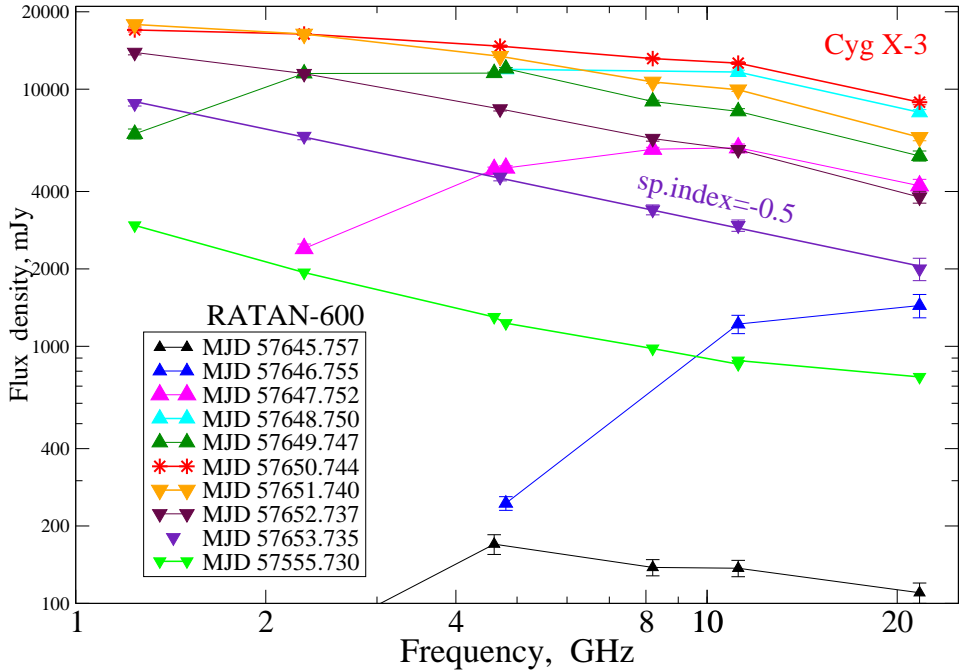
The radio spectra during first ten days of flare.
There is clear transition from the optically thick mode
to the optically thin one after MJD 57650.}

\subsection*{Discussion}

The giant flares have been often detected in the GBI two-frequency monitoring
program and \cite{Wa1} have  detected that flares occur after
'quenched state', when radio fluxes decreased to 10-30 mJy  at 2-8 GHz.
\cite{M1} analyzed the giant flare of 1999 and found that the radio fluxes
have anti-correlated with the hard (BATSE) X-ray fluxes and correlated during
the flare. The active period of the Cyg~X-3 in 2006-2009 showed similar
dependencies between soft (RXTE ASM), hard (Swift/BAT) X-rays and
radio emission (\cite{W1,C1,T06} or even with gamma-ray emission
\cite{Ta1,P1,P2}. The accretion disk-jet coupling in X-ray binaries
has been discussed during last 10-15 years especially in the frame of
the hardness-intensity diagram (HID) studies (\cite{F1}).
Based on the first-time developed HID of the microquasar Cyg~X-3 \cite{K1}
have detected the 'jet-line' of  the powerful ejections only after
so-called a 'hyper-soft' state, when hard X-ray fluxes fallen down to
detection level, meanwhile soft X-ray emission stays on high level.
\cite{T06} have successfully applied computer routine
to model radio flaring activity (in July 2006) of Cyg~X-3,
based on the model created by \cite{Ma1} and found main parameters:
magnetic field ($\sim 0.05$Gs), thermal electron densities
($3\times 10^5$)cm$^{-3}$ and the bulk speed of jets ($\sim 0.5$c).
The spectral evolution of the giant flare is described by a single
(during 3-4 days) ejection of the relativistic electrons, that moved
with high velocity ($\sim 0.5$c) away from the binary and expanded as
a conical structure. During  first days of the ejection jets is probably
optically thick due to synchrotron self-absorption or by thermal electrons
mixed with relativistic ones.  It is interesting that just in the beginning
of the new flare in September 2016 the MAXI sort X-ray (2-20 keV) fluxes
decreased from 0.35 crabs to 0.1 crabs thus Cyg~X-3 returned in hard state.

A lot of collected  measurements of the flare
with different  telescopes will be presented in the preparing paper.

\acknowledgements
S.A.T. acknowledges support through the Russian Government Program of
Competitive Growth of Kazan Federal University.



\end{document}